\documentclass[12pt,journal,draftcls,letterpaper,onecolumn]{IEEEtran}
\usepackage{amsmath,cite,graphicx,textcomp,url}

\newcommand{\bx}{\mathbf{x}}
\newcommand{\bt}{\mathbf{t}}
\begin{document}
\title{Extending the Recursive Jensen-Shannon Segmentation of
Biological Sequences}
\author{Siew-Ann Cheong,\thanks{S.-A. Cheong completed this work as a
Postdoctoral Associate with the Cornell Theory Center, Cornell
University, Ithaca, NY 14853.  He is presently an Assistant Professor
of Physics and Applied Physics with the School of Physical and
Mathematical Sciences, Nanyang Technological University, 21 Nanyang
Link, Singapore 637371, Republic of Singapore.  Email:
cheongsa@ntu.edu.sg.}
Paul
Stodghill,\thanks{P. Stodghill, D. J. Schneider and S. W. Cartinhour are with
the USDA Agricultural Research Service, Ithaca, NY 14853.  Email:
ps27@cornell.edu, djs30@cornell.edu, sc167@cornell.edu.} David J. Schneider,
Samuel W.  Cartinhour, and Christopher R. Myers\thanks{
C. R. Myers is with the Center for Advanced Computing, Cornell University, Ithaca, NY 14853.
Email: myers@tc.cornell.edu.} }
\maketitle

\begin{abstract}

In this paper, we extend a previously developed recursive entropic
segmentation scheme for applications to biological sequences.  Instead
of Bernoulli chains, we model the statistically stationary segments in
a biological sequence as Markov chains, and define a generalized
Jensen-Shannon divergence for distinguishing between two Markov
chains.  We then undertake a mean-field analysis, based on which we
identify pitfalls associated with the recursive Jensen-Shannon
segmentation scheme.  Following this, we explain the need for
segmentation optimization, and describe two local optimization schemes
for improving the positions of domain walls discovered at each
recursion stage.  We also develop a new termination criterion for
recursive Jensen-Shannon segmentation based on the strength of
statistical fluctuations up to a minimum statistically reliable
segment length, avoiding the need for unrealistic null and alternative
segment models of the target sequence.  Finally, we compare the
extended scheme against the original scheme by recursively segmenting
the \emph{Escherichia coli} K-12 MG1655 genome.

\end{abstract}


\section{Introduction}

\PARstart{L}{arge}-scale genomic rearrangements, such as
transpositions, inversions, and horizontal gene transfer (HGT), play
important roles in the evolution of bacteria.  Biological functions
can be lost or gained in such recombination events.  For example, it
is known that virulent genes are frequently found near the boundaries
of HGT islands, suggesting that virulence arise from the incorporation
of foreign genetic material
\cite{Hacker1997MolecularMicrobiology23p1089,
Groisman1997TrendsinMicrobiology5p343,
Hacker2000AnnRevMicrobiol54p641}.  In a recent essay, Goldenfeld and
Woese argued that the mosaic nature of bacterial genomes resulting
from such large-scale genomic rearrangements requires us to rethink
familiar notions of phylogeny and evolution
\cite{Goldenfeld2007Nature445p369}.  As a first step in unraveling the
complex sequence of events that shape the probable evolutionary
history of a bacterium, we need to first identify the recombination
sites bounding recombined segments, which are frequently
distinguishable statistically from their flanking sequences.

We do this by modeling the native and recombined segments in a genome
as stationary Markov chains.  The boundaries between such
statistically stationary segments (or domains) are called \emph{change
points} in the statistical modeling literature, or \emph{domain walls}
in the statistical physics literature.  Given a nucleotide or amino
acid sequence of length $N$, the problem of finding $M$ segments
generated by $P$ stationary Markov chains is called
\emph{segmentation} \cite{Carlstein1994ChangePointProblems,
Chen2000ParametricStatisticalChangePointAnalysis}.  Many segmentation
schemes can be found in the literature (see minireview by Braun and
Muller \cite{Braun1998StatisticalScience13p142}).  For $M \neq P$ both
unknown, Gionis and Mannila showed that finding the optimal
segmentation for a given sequence is NP-hard
\cite{Gionis2003RECOMB03p123}.  Therefore, some segmentation schemes
assume $P = M$, while others assume that $P$ is small, and known
beforehand.

Of these, the recursive segmentation scheme introduced by
Bernaola-Galv\'an \emph{et al.}
\cite{BernaolaGalvan1996PhysicalReviewE53p5181,
RomanRoldan1998PhysicalReviewLetters80p1344} is conceptually appealing
because of its simplicity.  In this scheme, a given sequence is
recursively partitioned into finer and finer segments --- all modeled as
Bernoulli chains --- based on their Jensen-Shannon divergences.  The
unknown number of segments $M$ (assumed to be equal to the number of
segment types $P$) is then discovered when segmentation is terminated
based on an appropriate statistical criterion.  In this paper, we
describe our extensions to this recursive segmentation scheme.  In
Sec.~\ref{section:generalizedJensenShannondivergences}, we explain how
Markov chains model the short-range correlations in a given sequence
better than Bernoulli chains, and thereafter generalize the
Jensen-Shannon divergence to distinguish between two Markov chains.
In Sec.~\ref{section:meanfieldanalysis}, we carry out a mean-field
analysis to better understand the recursive segmentation scheme and
its pitfalls, before describing two local segmentation optimization
algorithms for improving the statistical significance of domain walls
in Sec.~\ref{section:recursiveschemewithsegmentationoptimization}.  We
also develop in Sec.~\ref{section:segmentationterminationcondition} a
new termination criterion, based on the intrinsic statistical
fluctuations of the sequence to be segmented, for the recursive
segmentation scheme.  Finally, we compare our extended scheme against
the original scheme by recursively segmenting the \emph{Escherichia
coli} K-12 MG1655 genome in Sec.~\ref{subsection:recursivegenome},
before concluding in Sec.~\ref{section:conclusions}.

\section{Modeling Segments As Markov Chains}
\label{section:generalizedJensenShannondivergences}

In the earliest recursive segmentation scheme proposed by
Bernaola-Galv\'an \emph{et al.}
\cite{BernaolaGalvan1996PhysicalReviewE53p5181,
RomanRoldan1998PhysicalReviewLetters80p1344}, the divergence between
1-mer statistics from two or more subsequences of a given sequence is
examined.  These subsequences are modeled as Bernoulli chains
(equivalent to Markov chains of order $K = 0$), even though it is well
known that biological sequences exhibit dinucleotide correlations and
codon biases \cite{Grantham1981NucleicAcidsResearch9pR43,
Shepherd1981ProcNatlAcadSciUSA78p1596,
Staden1982NucleicAcidsResearch10p141,
Fickett1982NucleicAcidsResearch10p5303, Herzel1995PhysicaA216p518}.
Later versions of the recursive segmentation scheme examine higher
order subsequence statistics, so as to take advantage of different
codon usage in coding and noncoding regions
\cite{BernaolaGalvan2000PhysicalReviewLetters85p1342,
Nicorici2003FINSIG03,
Nicorici2004EURASIPJournalonAppliedSignalProcessing1p81}, but these
are still assumed to be drawn from Bernoulli chains, albeit with
extended alphabets.  The first study we are aware of modeling
subsequences as Markov chains for recursive segmentation is the work
by Thakur \emph{et al.} \cite{Thakur2007PhysicalReviewE75a011915}.

In this section, we will explain why the observed dinucleotide
frequences and codon biases in biological sequences can be better
modeled by Markov chains of order $K > 0$, compared to Bernoulli
chains with the same high order statistics.  We will then generalize
the Jensen-Shannon divergence, so that it can be used in entropic
segmentation schemes to quantify the statistical difference between
Markov chains of order $K > 0$. Finally, we discuss the added modeling
complexities associated with using Markov-chain orders that vary from
segment to segment, and change when segments are further divided.

\subsection{Markov Chains Versus Bernoulli Chains}

Given a sequence $\mathbf{x} = x_1 x_2 \cdots x_N$, where the symbols
$x_i$ are drawn from an alphabet $\mathcal{S} = \{\alpha_s\}_{s=1}^S$
containing $S$ letters, we want to model $\bx$ as being generated
sequentially from a single stationary stochastic process.  In the
bioinformatics literature, $\mathbf{x}$ is usually modeled as a
Bernoulli chain or as a Markov chain.  For a Bernoulli chain, the $N$
symbols are obtained from $N$ independent trials, governed by the
\emph{state probabilities}
\begin{equation}
P(x_i = \alpha_s) = P(\alpha_s) = P_s,
\end{equation}
whereas for a Markov chain of order $K$, the probability 
\begin{equation}
p(x_i = \alpha_s | x_{i-1} = \alpha_{t_1}, \dots, x_{i-K} = \alpha_{t_K}) 
\end{equation}
of finding the $i$th symbol to be $x_i = \alpha_s$ is conditioned by
the $K$ symbols preceding it.  We call these the \emph{transition
probabilities}
\begin{equation}
p(\alpha_s | \alpha_{t_1} \cdots \alpha_{t_K}) = p_{\mathbf{t}s}
\end{equation}
of going from the $K$-mer $\alpha_{t_1}\cdots\alpha_{t_K}$ to the
1-mer $\alpha_s$.  For the rest of this paper, we use the shorthand
$\mathbf{t} \to s$ to represent the transition
$\alpha_{t_1}\cdots\alpha_{t_K} \to \alpha_s$.

A Bernoulli chain over the alphabet $\mathcal{S}$ is equivalent to a
Markov chain of order $K = 0$, and is completely uncorrelated, in the
sense that $P(x_i, x_j) = P(x_i) P(x_j)$.  Markov chains of order $K >
0$, on the other hand, contain short-range correlations, the scale of
which is set by $K$, but no long-range correlations.  Therefore, if
the target sequence $\mathbf{x}$ contains short-range correlations, a
Markov chain of nonzero order models $\mathbf{x}$ better than a
Bernoulli chain over the the same alphabet.  It is also possible to
capture short-range correlations in $\mathbf{x}$ using Bernoulli
chains over extended alphabets.  For example, a Markov chain of order
$K$ over $\mathcal{S}$ and a Bernoulli chain over the extended
alphabet $\mathcal{S}^{K+1}$ can both be used to model the $(K+1)$-mer
statistics of $\mathbf{x}$.  They are, however, not equivalent.

Let $f_{\mathbf{t}s}$ be the number of times the $(K+1)$-mer
$\alpha_{t_K}\cdots\alpha_{t_1}\alpha_s$ appears in $\mathbf{x}$.  To
model $\mathbf{x}$ as a Markov chain of order $K$ over $\mathcal{S}$,
we must use these $(K+1)$-mer counts to make maximum-likelihood
estimates
\begin{equation}\label{equation:maxlikelihoodtransitionprobabilities}
\hat{p}_{\mathbf{t}s} = \frac{f_{\mathbf{t}s}}{\sum_{s' = 1}^S
f_{\mathbf{t}s'}}
\end{equation}
of the $S^K(S - 1)$ independent transition probabilities, subject to
the normalization
\begin{equation}
\sum_{\mathbf{t} \in \mathcal{S}^{K}}\sum_{s=1}^S f_{\mathbf{t}s} = N,
\end{equation}
i.e. every sequence position in $\mathbf{x}$ contributes one count to
the model estimation.  A Bernoulli chain over the extended alphabet
$\mathcal{S}^{K+1}$, on the other hand, contains $(S^{K+1} - 1)$
independent state probabilities --- $(S^K - 1)$ independent parameters
more than the Markov chain of order $K$.

Besides having many more independent parameters, there is also the
subtle question of how to estimate the state probabilities, if we are
to model $\mathbf{x}$ using the Bernoulli chain over
$\mathcal{S}^{K+1}$.  If the state probabilities $\{P_{\mathbf{t}s}\}$
were given, we would generate a Bernoulli chain by sequentially
appending $(K+1)$-mers drawn according to $\{P_{\mathbf{t}s}\}$.  This
suggests that we partition the observed sequence $\mathbf{x}$ into
nonoverlapping $(K+1)$-mers, and use their counts $f'_{\mathbf{t}s}$
to determine the maximum-likelihood state probabilities
\begin{equation}
\hat{P}_{\mathbf{t}s} = \frac{f'_{\mathbf{t}s}}
{\sum_{\mathbf{t}' \in \mathcal{S}^{K}}\sum_{s'=1}^S f'_{\mathbf{t}' s'}}.
\end{equation}
However, only one in $(K + 1)$ sequence positions in $\mathbf{x}$
contributes a count to the model estimation.  The result is that such
a Bernoulli chain model of $\mathbf{x}$ would be much less
statistically significant than a order-$K$ Markov chain model of
$\mathbf{x}$, since we are using a smaller number of counts to
estimate a larger number of parameters.  Alternatively, we can note
the fact that there are $(K+1)$ different ways to partition
$\mathbf{x}$ into nonoverlapping $(K+1)$-mers.  If we combine the
counts from these different partitions, then every sequence position
contributes one count to model estimation, just as for the Markov
chain model.

No matter how we perform the model estimation, two adjacent
$(K+1)$-mers in a Bernoulli chain over $\mathcal{S}^{K+1}$ are
guaranteed to be uncorrelated.  This means that correlations at the
$(K+1)$-mer level in $\mathbf{x}$ can only be partly captured by a
maximum-likelihood Bernoulli chain over $\mathcal{S}^{K+1}$.  In
contrast, a maximum-likelihood Markov chain of order $K$ over
$\mathcal{S}$ will capture most of the $(K+1)$-mer correlations in
$\mathbf{x}$.  We can ensure that most or all of the $(K+1)$-mer
correlations in $\mathbf{x}$ are captured by a Bernoulli chain model,
by going to even larger extended alphabets, but as we have seen above,
the statistical quality of the model deteriorates rapidly as we try to
model $N$ counts with an exponentially increasing numbers of
parameters.  Based on the discussions above comparing Markov chain and
Bernoulli chain models, we argue that a Markov chain of order $K$ is a
more compact model of the $(K+1)$-mer statistics of an observed
sequence $\mathbf{x}$.

\subsection{Generalizing the Jensen-Shannon Divergence}

After the maximum-likelihood model of an observed sequence
$\mathbf{x}$ is determined, we can compute its \emph{sequence
likelihood} within the model.  This is the probability of getting
$\mathbf{x}$, if we use the maximum-likelihood model to generate
random sequences of length $N$.  For a Bernoulli chain model of
$\mathbf{x}$ over $\mathcal{S}$, we find the sequence likelihood to be
given by
\begin{equation}
P_1(\mathbf{x}) = \prod_{s=1}^S \left(\hat{P}_s\right)^{f_s},
\end{equation}
where $f_s$ is the number of times the 1-mer $\alpha_s$ appears in
$\mathbf{x}$.

Now, if we suspect that $\mathbf{x}$ actually comprises two
statistically stationary segments $\mathbf{x}_L = x_1 x_2 \cdots x_i$
and $\mathbf{x}_R = x_{i+1} x_{i+2} \cdots x_N$, with domain wall at
the \emph{cursor position} $i$, we must determine one
maximum-likelihood Bernoulli chain model for $\mathbf{x}_L$, and
another for $\mathbf{x}_R$.  This involves tallying the 1-mer counts
$f_s^L$ and $f_s^R$ for $\mathbf{x}_L$ and $\mathbf{x}_R$
respectively, and then estimating the maximum-likelihood state
probabilities
\begin{equation}
\hat{P}_s^L = \frac{f_s^L}{\sum_{s'=1}^S f_{s'}^L}, \quad
\hat{P}_s^R = \frac{f_s^R}{\sum_{s'=1}^S f_{s'}^R}.
\end{equation}
In this two-Bernoulli-segment model of $\mathbf{x}$, the sequence
likelihood is given by
\begin{equation}
P_2(\mathbf{x}, i) = \prod_{s=1}^S
\left(\hat{P}_s^L\right)^{f_s^L}
\left(\hat{P}_s^R\right)^{f_s^R}.
\end{equation}
Because we have more free parameters in the two-segment model to fit
1-mer statistics in the observed sequence $\mathbf{x}$, we have
$P_2(i) \geq P_1$ for all $i$.  The Jensen-Shannon divergence
\cite{Lin1991IEEETransactionsonInformationTheory37p145}
\begin{equation}\label{equation:JensenShannonBernoulli}
\Delta(i) = \log\frac{P_2(i)}{P_1} 
= \sum_{s=1}^S \left[-f_s \log \hat{P}_s +
f_s^L \log \hat{P}_s^L +
f_s^R \log \hat{P}_s^R\right] \geq 0,
\end{equation}
a symmetric variant of the relative entropy known more commonly as the
\emph{Kullback-Leibler divergence}, is then a quantitative measure of
how much better the two-segment model fits $\mathbf{x}$, compared to
the one-segment model, subject to the constraints
\begin{equation}
f_s^L + f_s^R = f_s, \quad
\sum_{s=1}^S f_s = N.
\end{equation}
In the literature, the Jensen-Shannon divergence is only defined for
proper distributions $\{\hat{P}_s^L\}$, $\{\hat{P}_s^R\}$, and
$\{\hat{P}_s\}$, for which
\begin{equation}
\sum_{s=1}^S \hat{P}_s^L = \sum_{s=1}^S \hat{P}_s^R = 
\sum_{s=1}^S \hat{P}_s = 1.
\end{equation}
When we want to compare the $(K+1)$-mer
statistics of the subsequences $\bx_L$ and $\bx_R$, by modeling these
as Bernoulli chains, the expression for the Jensen-Shannon divergence
becomes
\begin{equation}
\Delta(i) = \sum_{\bt \in \mathcal{S}^{K}}\sum_{s=1}^S 
\left[-f_{\bt s} \log \hat{P}_{\bt s} +
f_{\bt s}^L \log \hat{P}_{\bt s}^L +
f_{\bt s}^R \log \hat{P}_{\bt s}^R\right],
\end{equation}
comparing the proper distributions $\{\hat{P}_{\bt s}^L\}$ and
$\{\hat{P}_{\bt s}^R\}$ against $\{\hat{P}_{\bt s}\}$, which satisfy
the normalization condition
\begin{equation}
\sum_{\bt \in \mathcal{S}^{K}}\sum_{s=1}^S \hat{P}_{\bt s}^L = 
\sum_{\bt \in \mathcal{S}^{K}}\sum_{s=1}^S \hat{P}_{\bt s}^R =
\sum_{\bt \in \mathcal{S}^{K}}\sum_{s=1}^S \hat{P}_{\bt s} = 1.
\end{equation}

As we have explained, the $(K+1)$-mer statistics of $\bx_L$, $\bx_R$,
and $\bx$ are better modeled as order-$K$ Markov chains over
$\mathcal{S}$.  These maximum-likelihood one-segment and two-segment
Markov-chain models are determined by tallyng the transition counts
$f_{\mathbf{t}s}$, $f_{\mathbf{t}s}^L$, and $f_{\mathbf{t}s}^R$, and
estimating the transition probabilities $\hat{p}_{\mathbf{t}s}$,
$\hat{p}_{\mathbf{t}s}^L$, and $\hat{p}_{\mathbf{t}s}^R$ according to
Eq.~\eqref{equation:maxlikelihoodtransitionprobabilities}.  However,
the full sets $\{\hat{p}_{\bt s}^L\}$, $\{\hat{p}_{\bt s}^R\}$, and
$\{\hat{p}_{\bt s}\}$ of transition probabilities are not proper
distributions, because they sum to 
\begin{equation}
\sum_{\bt \in \mathcal{S}^K}\sum_{s=1}^S \hat{p}_{\bt s}^L =
\sum_{\bt \in \mathcal{S}^K}\sum_{s=1}^S \hat{p}_{\bt s}^R = 
\sum_{\bt \in \mathcal{S}^K}\sum_{s=1}^S \hat{p}_{\bt s} = S^K.
\end{equation}
Rather, we must think of them as $S^K$ sets of proper
distributions, in which case the Jensen-Shannon divergence that
simultaneously compares all $S^K$ distributions generalizes to 
\begin{equation}\label{equation:JensenShannondivergence}
\begin{split}
\Delta(i) &= \sum_{\mathbf{t} \in \mathcal{S}^{K}}\sum_{s = 1}^S 
\left[-f_{\mathbf{t}s} \log \hat{p}_{\mathbf{t}s} + 
f_{\mathbf{t}s}^L \log \hat{p}_{\mathbf{t}s}^L + 
f_{\mathbf{t}s}^R \log \hat{p}_{\mathbf{t}s}^R \right].
\end{split}
\end{equation}
A similar expression for $\Delta(i)$ was derived by Thakur \emph{et
al.} \cite{Thakur2007PhysicalReviewE75a011915}.

Just as for Bernoulli-chain models, the Jensen-Shannon divergence in
Eq.~\eqref{equation:JensenShannondivergence} for Markov-chain models
of $\bx_L$, $\bx_R$, and $\bx$, is also the log ratio of the
one-segment and two-segment sequence likelihoods
\begin{equation}\label{equation:Markovchainsequencelikelihood}
P_1(\mathbf{x}) = \prod_{\mathbf{t} \in \mathcal{S}^{K}} \prod_{s=1}^S 
\left(\hat{p}_{\mathbf{t}s}\right)^{f_{\mathbf{t}s}}, \quad
P_2(\mathbf{x}, i) = \prod_{\mathbf{t} \in \mathcal{S}^{K}} \prod_{s=1}^S 
\left(\hat{p}_{\mathbf{t}s}^L\right)^{f_{\mathbf{t}s}^L}
\left(\hat{p}_{\mathbf{t}s}^R\right)^{f_{\mathbf{t}s}^R},
\end{equation}
respectively.  Here let us note that the summary of transition counts
$\{f_{\mathbf{t}s}, f_{\mathbf{t}s}^L, f_{\mathbf{t}s}^R\}$ is
satisfied by more than one sequence, but all these sequences have the
same sequence likelihoods within the maximum-likelihood Markov chain
models.

\subsection{Using Variable Markov-Chain Orders for Segmentation}
\label{section:modelselection}

When all the segments in a sequence are modeled as Bernoulli chains,
we need to determine only the optimal domain wall positions $\{i_1,
i_2, \dots, i_M\}$.  When the segments are modeled as Markov chains,
we need also decide what Markov-chain orders $\{K_1, K_2, \dots,
K_{M+1}\}$ to use for the segments.  These two problems are coupled,
and we shall understand in this subsection how to solve them, within
the recursive segmentation framework, by considering the refinement of
a sequence into two segments.

We start by considering the one-segment Markov-chain model of the
sequence $\mathbf{x}$, which serves as the null model against which
the two-segment Markov-chains model is compared.  In principle, we can
fit $\mathbf{x}$ to a one-segment Markov-chain model of any order $K$.
The one-segment sequence likelihood $P_1(K)$ given in
Eq.~\eqref{equation:Markovchainsequencelikelihood} increases with $K$,
because of the exponentially increasing number of free parameters to
fit the observed counts.  To perform segmentation, we want $K$ to be
as large as possible, so that we can distinguish segments differing
only in their high-order statistics. To determine the maximum
Markov-chain order $K^*$ justified by the statistics of $\mathbf{x}$,
we compare the \emph{penalized sequence likelihoods} for various $K$.
The negative logarithms of these penalized sequence likelihoods are
also called the \emph{information criteria}
\begin{equation}
\phi = -2\log P_1(K) + \theta,
\end{equation}
where $\theta$ is a penalty function.  Common information criteria
found in the literature are, the Akaike information criterion (AIC)
\cite{Akaike1974IEEETransactionsAutomaticControl19p716}, the Schwarz
information criterion (SIC) \cite{Schwarz1978AnnalsofStatistics6p461},
and the Bayes information criterion (BIC)
\cite{Katz1981Technometrics23p243}.  For a Markov chain of length $N$,
order $K$ over an alphabet of $S$ letters, these differ in their
penalty functions
\begin{equation}
\theta(N, S, K) = \begin{cases}
S^K (S - 1), & \text{AIC}; \\
\frac{1}{2} S^{K+1} \log N, & \text{SIC}; \\
S^K (S - 1) \log N, & \text{BIC}.
\end{cases}
\end{equation}
In particular, Katz showed that the BIC gives an unbiased estimate of
the true order of a Markov chain \cite{Katz1981Technometrics23p243},
so we use the Markov-chain order $K^*$ minimizing the BIC as the order
of our one-segment model of $\mathbf{x}$.

Since we do not know beforehand whether $\mathbf{x}$ contain
statistically distinct segments, or whether the segments, if present,
differ in low-order or high-order statistics, it is safest to search
for them at $K^*$, the maximum statistically justifiable Markov-chain
order.  In the recursive segmentation scheme described in this paper,
we would compute the Jensen-Shannon divergence $\Delta(i)$ given in
Eq.~\eqref{equation:JensenShannondivergence} as a function of the
cursor position $i$ at order $K^*$, and partition $\mathbf{x}$ into
two segments $\mathbf{x}_L \equiv x_1 \dots x_{i^*}$ and $\mathbf{x}_R
\equiv x_{i^*+1} \cdots x_N$, where $i^*$ is the sequence position
maximizing the order-$K^*$ Jensen-Shannon divergence.  Once the
segments are discovered, we can compute the maximum Markov-chain
orders $K_L^*$ and $K_R^*$ that we would use to further partition
$\mathbf{x}_L$ and $\mathbf{x}_R$ by minimizing their respective BICs.
Because $\mathbf{x}_L$ and $\mathbf{x}_R$ are shorter than
$\mathbf{x}$, we naturally have $K_L^*, K_R^* \leq K^*$.

Next, suppose we believe that $\mathbf{x}$ is indeed composed of two
statistically distinct segments, $\mathbf{x}_L$ and $\mathbf{x}_R$,
but $i^*$ is not the best position for the domain wall between them.
To find a better position $i^{**}$ for the domain wall between
$\mathbf{x}_L$ and $\mathbf{x}_R$, we can compute the two-segment
sequence likelihoods $P_2(i)$ for $\mathbf{x}$ at various cursor
positions $i$, using an order-$K_L^*$ model for $\mathbf{x}_L$, and an
order-$K_R^*$ model for $\mathbf{x}_R$, and pick $i^{**}$ to be the
sequence position maximizing $P_2$.  Alternatively, we can compute the
Jensen-Shannon divergence spectrum $\Delta(i)$ at Markov-chain order
$K = \min(K_L^*, K_R^*)$, and accept as $i^{**}$ its divergence
maximum.  We choose $K = \min(K_L^*, K_R^*)$ for doing so because this
Markov-chain order is statistically justifiable in both segments, and
also has a smaller uncertainty associated with the domain wall
position, as discussed in
Sec.~\ref{section:disproportionatecontributionbyrarestates}.

\section{Mean-Field Picture of Recursive Jensen-Shannon Segmentation}
\label{section:meanfieldanalysis}

In statistical physics, intrinsic fluctuations in the properties of a
physical system (for example, the local density in a fluid) makes its
true behaviour difficult to analyze and understand.  In many physical
systems, however, a great deal about their properties can be
understood from simplified mean-field pictures, where we ignore
statistical fluctuations, and assume that these properties take on
system-wide values.  In the same spirit, we develop in this section a
mean-field picture of the recursive Jensen-Shannon segmentation scheme
proposed by Bernaola-Galv\'an \emph{et al.}
\cite{BernaolaGalvan1996PhysicalReviewE53p5181,
RomanRoldan1998PhysicalReviewLetters80p1344}, and explain the need to
move or remove domain walls that have lost statistical significance as
the segmentation is recursively refined.  We start by examining in
Sec.~\ref{section:disproportionatecontributionbyrarestates} the
general anatomy of a Jensen-Shannon divergence spectrum, and how
transitions rare on one side of the cursor position give rise to
`noise' in the spectrum that cloud our understanding of recursive
segmentation.  We argue that a mean-field picture will be helpful, and
thus proceed in Sec.~\ref{subsection:meanfieldwindowlessspectrum}
to define the appropriate mean-field limit, and thereafter perform
mean-field analysis on the recursive Jensen-Shannon segmentation
scheme.

\subsection{Influence of Rare Transitions on the Jensen-Shannon
Divergence Spectrum}
\label{section:disproportionatecontributionbyrarestates}

As discussed earlier, the most straightforward way to determine the
optimal point $i^*$ to cut a given sequence $\mathbf{x} = x_1 x_2
\cdots x_N$ into two segments is to compute the Jensen-Shannon
divergence spectrum $\Delta(i)$ between the left segment $\mathbf{x}_L
= x_1 x_2 \cdots x_i$ and the right segment $\mathbf{x}_R = x_{i+1}
x_{i+2} \cdots x_N$ for all cursor positions $0 < i < N$, and then
determine the divergence maximum $i^*$ such that $\Delta(i^*) = \max_i
\Delta(i)$.  As we move away from $i^*$, each step we take results in
the left and right distributions increasing or decreasing by one
transition, giving rise to discrete jumps $\delta\Delta$ in the
Jensen-Shannon divergence.  From the definition of the Jensen-Shannon
divergence in Eq.~\eqref{equation:JensenShannondivergence}, we find that
\begin{equation}\label{eqn:jumpA}
\delta\Delta = 
\sum_{\mathbf{t} \in \mathcal{S}^{K}}\sum_{s = 1}^S
\left[\delta f_{\mathbf{t}s}^L \log \hat{p}_{\mathbf{t}s}^L +
\frac{f_{\mathbf{t}s}^L}{\hat{p}_{\mathbf{t}s}^L}\, 
\delta \hat{p}_{\mathbf{t}s}^L \right] + 
\sum_{\mathbf{t} \in \mathcal{S}^{K}}\sum_{s = 1}^S
\left[\delta f_{\mathbf{t}s}^R \log \hat{p}_{\mathbf{t}s}^R +
\frac{f_{\mathbf{t}s}^R}{\hat{p}_{\mathbf{t}s}^R}\, 
\delta \hat{p}_{\mathbf{t}s}^R \right],
\end{equation}
for some given changes $\delta f_{\mathbf{t}s}^L = -\delta
f_{\mathbf{t}s}^R$ to the transition counts.  When the cursor position
is shifted over by one base, only a single transition $\mathbf{t} \to s$
is affected, for which $\delta f_{\mathbf{t}s}^L = \pm 1 = -\delta
f_{\mathbf{t}s}^R$.  However, all $S$ transition probabilities
associated with the $K$-mer $\alpha_{t_1}\alpha_{t_2} \cdots
\alpha_{t_K}$ are affected, and we have to write the jump as
\begin{equation}\label{eqn:jumpB}
\delta\Delta = 
\sum_{s'=1}^S \left[\delta f_{\mathbf{t}s'}^L \log \hat{p}_{\mathbf{t}s'}^L + 
\frac{f_{\mathbf{t}s'}^L}{\hat{p}_{\mathbf{t}s'}^L}\, 
\delta \hat{p}_{\mathbf{t}s'}^L \right] + 
\sum_{s'=1}^S \left[\delta f_{\mathbf{t}s'}^R \log \hat{p}_{\mathbf{t}s'}^R + 
\frac{f_{\mathbf{t}s'}^R}{\hat{p}_{\mathbf{t}s'}^R}\, 
\delta \hat{p}_{\mathbf{t}s'}^R \right],
\end{equation}
where we demand that $\delta f_{\mathbf{t}s'}^L = \delta_{ss'} =
-\delta f_{\mathbf{t}s'}^R$.  

Noting that
\begin{equation}\label{eqn:jumpC}
\hat{p}_{\mathbf{t}s}^{L, R} = 
\frac{f_{\mathbf{t}s}^{L, R}}{f_{\mathbf{t}}^{L, R}}, \quad
f_{\mathbf{t}}^{L, R} = \sum_{s'} f_{\mathbf{t}s'}^{L, R}, \quad
\delta \hat{p}_{\mathbf{t}s}^{L, R} = \frac{\delta f_{\mathbf{t}s}^{L,
R}}{f_{\mathbf{t}}^{L, R}} - \frac{f_{\mathbf{t}s}^{L,
R}}{\left(f_{\mathbf{t}}^{L, R}\right)^2}\, \delta f_{\mathbf{t}}^{L, R},
\end{equation}
and also
\begin{equation}\label{eqn:jumpE}
\frac{f_{\mathbf{t}s}^{L, R}}{\hat{p}_{\mathbf{t}s}^{L, R}}\, \delta
\hat{p}_{\mathbf{t}s}^{L, R} = f_{\mathbf{t}}^{L, R}\, \delta
\hat{p}_{\mathbf{t}s}^{L, R} 
= \delta f_{\mathbf{t}s}^{L, R} -
\frac{f_{\mathbf{t}s}^{L, R}}{f_{\mathbf{t}}^{L, R}}\, 
\delta f_{\mathbf{t}}^{L, R} 
= \delta f_{\mathbf{t}s}^{L, R} -
\hat{p}_{\mathbf{t}s}^{L, R}\, \delta f_{\mathbf{t}}^{L, R},
\end{equation}
we simplify the expression for the divergence jump to
\begin{equation}
\begin{aligned}
\delta\Delta &= 
\sum_{s'=1}^S \delta f_{\mathbf{t}s'}^L \log \hat{p}_{\mathbf{t}s'}^L +
\sum_{s'=1}^S \delta f_{\mathbf{t}s'}^L - 
\sum_{s'=1}^S \hat{p}_{\mathbf{t}s'}^L\, \delta f_{\mathbf{t}}^L +
\sum_{s'=1}^S \delta f_{\mathbf{t}s'}^R \log \hat{p}_{\mathbf{t}s'}^R
+ {} \\
&\quad\
\sum_{s'=1}^S \delta f_{\mathbf{t}s'}^R - \sum_{s'=1}^S
\hat{p}_{\mathbf{t}s'}^R\, \delta f_{\mathbf{t}}^R \\
&= \log\frac{\hat{p}_{\mathbf{t}s}^L}{\hat{p}_{\mathbf{t}s}^R},
\end{aligned}
\end{equation}
where we made use of the facts that $\delta f_{\mathbf{t}s}^L =
-f_{\mathbf{t}s}^R$, $\delta f_{\mathbf{t}}^L = -\delta
f_{\mathbf{t}}^R$, and $\sum_{s'=1}^S \hat{p}_{\mathbf{t}s}^{L, R} =
1$.  As we can see, for a unit shift to the right, the sign and
magnitude of the jump depends only on the transition probabilities,
but not the transition counts.  Here let us distinguish between common
and rare \emph{states}, versus common and rare \emph{transitions}.
Common (or rare) states are $(K + 1)$-mers with high (or low) counts
$f_{\mathbf{t}s}^{L, R}$, while common (or rare) transitions are $(K +
1)$-mers associated with high (or low) transition probabilities
$\hat{p}_{\mathbf{t}s}^{L, R}$.

\begin{figure}[htbp]
\centering
\includegraphics[scale=0.45,clip=true]{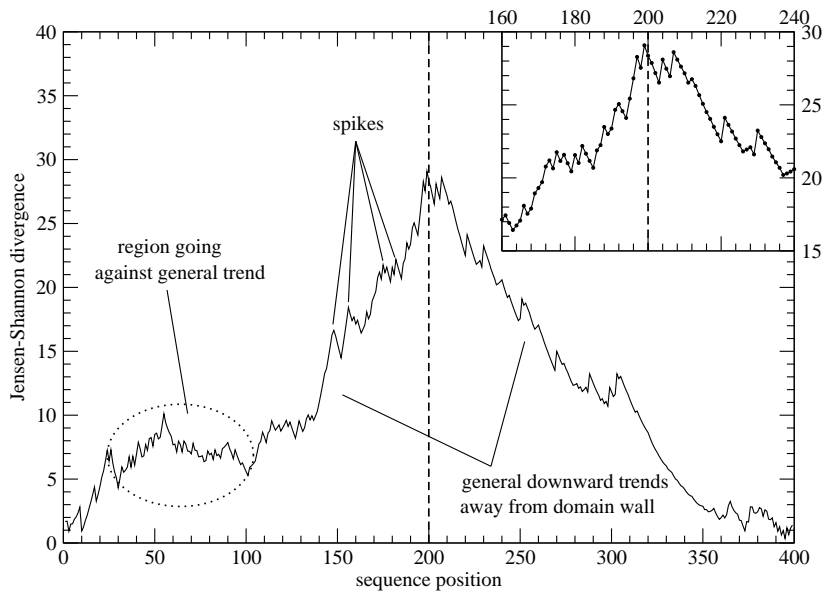}
\caption{A typical Jensen-Shannon divergence spectrum, showing the
general downward trends away from the domain wall (dashed vertical
line), along with spikes that arise from the discrete nature of the
sequence.  Sometimes, regions which go against the general trend will
also appear.  (Inset) A typical spike structure around the divergence
maximum.}
\label{figure:jstypical}
\end{figure}

Fig.~\ref{figure:jstypical} shows a typical Jensen-Shannon
divergence spectrum obtained from an artificial sequence of length $N
= 400$ over $S = 4$ letters, consisting of two length $N_L = N_R =
200$, $K = 0$ segments.  General downward trends left and right of the
single domain wall at $i = 200$ are observed in the Jensen-Shannon
divergence spectrum, though we sometimes find regions going against
the trend locally.  When this happens, it is statistically acceptable,
and frequently desirable, to place additional domain walls in the
sequence, even though in this example we know there is only one true
domain wall.  Spikes, which are large divergence jumps resulting from
transitions rare in one of the segments, feature ubiquitously in the
Jensen-Shannon divergence spectra of discrete sequences.

From this analysis, we learned that the uncertainty in determining the
true domain wall position is dictated by the \emph{competition}
between common and rare transitions.  The common transitions, which
are present in larger numbers, determine the average jumps
$\overline{\delta\Delta}_L$ and $\overline{\delta\Delta}_R$ and hence
the general downward trends left and right of the domain wall.  The
rare transitions, on the other hand, are associated with spikes (local
maxima) in the divergence spectrum.  Specifically, rare transitions
that are most asymmetrically distributed between the two segments are
the most important, since they give rise to the largest spikes
$|\delta\Delta|_{\max}$.  Moving $k$ bases away from the true domain
wall, we expect the Jensen-Shannon divergence to decrease by roughly
$k|\overline{\delta\Delta}|$.  If $k$ is small,
$k|\overline{\delta\Delta}| < |\delta\Delta|_{\max}$, and a single
maximal spike encountered within these $k$ bases will bring the
divergence up to a value greater than that at the true domain wall,
throwing off the cut that we make.  In contrast, a maximal spike
encountered after we have moved more than $k = |\delta\Delta|_{\max} /
|\overline{\delta\Delta}|$ bases away from the true domain wall will
not be able to raise the divergence beyond that observed at the true
domain wall.  The ratio $|\delta\Delta|_{\max} /
|\overline{\delta\Delta}|$ therefore gives a quantitative measure of
the uncertainty involved in determining the position of the true
domain wall.

For a fixed sequence length $N$, the number of rare transitions
increases, while the transition probabilities of the rarest
transitions decrease, with increasing $K$.  We thus find more and
stronger spikes.  For a fixed Markov-chain order $K$, the number of
rare transitions remain more or less constant with decreasing $N$, but
the transition probabilities of the rarest transitions decrease with
decreasing $N$ as a result of stronger statistical fluctuations in the
transition counts.  We therefore find stronger spikes.  The
proliferation of strong spikes makes segmentation unreliable, and also
distracts from our understanding of the recursive segmentation scheme.
This is where a mean-field picture, within which we can study the
progress of recursive segmentation in the absence of such statistical
fluctuations, would be extremely helpful.

\subsection{Jensen-Shannon Divergence Spectrum in the Mean-Field Limit}
\label{subsection:meanfieldwindowlessspectrum}

\begin{figure}[htbp]
\centering
\includegraphics[scale=0.2]{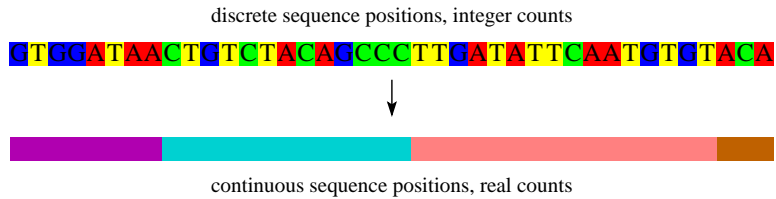}
\caption{Going from a discrete description to a continuum description
of a nucleotide sequence.}
\label{figure:meanfieldpicture}
\end{figure}

In a discrete genomic sequence of nucleotides, the sequence positions
$i$ and $j$ take on integer values.  The frequencies of various
$K$-mers occurring within the interval $[i, j > i)$ are also integers.
From the previous subsection, we understood how a spike in the
Jensen-Shannon divergence spectrum arise when there is a unit
increment for a rare transition count.  Such rare transitions, of
course, occur infrequently along the sequence.  If we can distribute
the unit increment associated with a rare transition over the interval
between two such transitions, the statistical effect of the increment
will not be piled up over a single base as a spike.  This can be done
not just for rare transitions, but for all transitions.  A continuum
description of the sequence can then be obtained by allowing both the
sequence positions and statistical frequencies to vary continuously,
as shown in Fig.~\ref{figure:meanfieldpicture}.

Depending on how we break the unit increments over one base into
fractional increments over many bases, there are many ways to
redistribute the transition counts over the interval $[i, j > i)$ of
the sequence.  If our goal is to model this interval as a stationary
Markov chain of order $K$, then in the \emph{mean-field limit} we
distribute the $(K+1)$-mer statistics within $[i, j)$ uniformly along
the interval.  In the mean-field limit so defined for $[i, j)$, the
count $f_{\mathbf{t}s}^{[i', j')}$ of the transition $\mathbf{t} \to
s$ within the subinterval $[i', j' > i') \subseteq [i, j)$ is given by
\begin{equation}
f_{\mathbf{t}s}^{[i', j')} =
\frac{j' - i'}{j - i}\,
f_{\mathbf{t}s}^{[i, j)},
\end{equation}
where $f_{\mathbf{t}s}^{[i, j)}$ is the net $\mathbf{t} \to s$
transition count within $[i, j)$.  In this way, we remove local
fluctuations in the $(K+1)$-mer statistics within the interval.

\begin{figure}[htbp]
\centering
\includegraphics[scale=0.45,clip=true]{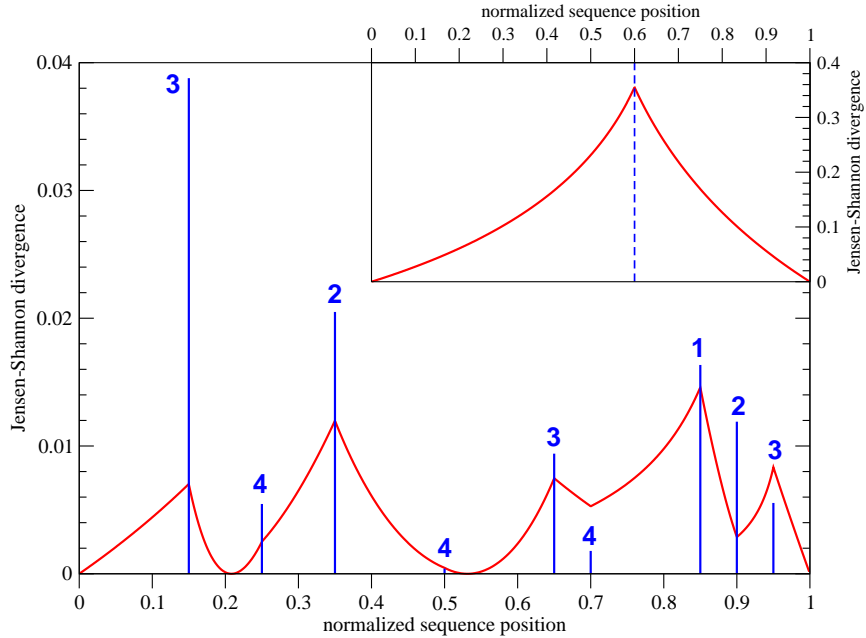}
\caption{The Jensen-Shannon divergence $\Delta(z)$ (red solid curve)
as a function of the normalized cursor position $z$ within an
artificial binary sequence composed of ten mean-field segments,
characterized by the probabilities (left to right) $\mathbf{P}(0) =
(0.55, 0.05, 0.20, 0.60, 0.65, 0.30, 0.45, 0.05, 0.45, 0.15)$.  The
blue bars indicate the strength of each of the nine domain walls,
while the number at each domain wall indicate which recursion step it
is discovered. (Inset) The Jensen-Shannon divergence $\Delta(z)$ (red
solid curve) as a function of the normalized cursor position $z$
within an artificial binary sequence composed of two mean-field
segments, characterized by the probabilities $P_L(0) = 0.1$ and
$P_R(0) = 0.9$.  The domain wall at $z = 0.60$ is indicated by the
blue dashed vertical line.}
\label{figure:JS10}
\end{figure}

In Fig.~\ref{figure:JS10}, we show as an example the Jensen-Shannon
divergence spectrum for an artificial binary sequence consisting of
ten mean-field segments.  As we can see, peaks in the divergence
spectrum occur only at domain walls, but not all domain walls appear
as peaks in the divergence spectrum.  Some domain walls manifest
themselves as kinks in the divergence spectrum, while others, under
special distributions of the segment statistics, may even have
vanishing divergences.  Performing recursive Jensen-Shannon
segmentation on this ten-segment sequence, we recover all nine domain
walls.  These, however, are not discovered in the order of their true
strengths (heights of the blue bars), measured by the Jensen-Shannon
divergence of the pairs of segments they separate.  For the example
shown, the third strongest domain wall is discovered in the first
recursion step, the second and fourth strongest domain walls in the
second recursion step, and the strongest domain wall discovered only
in the third recursion step.

\section{Optimized Recursive Segmentation Scheme}
\label{section:recursiveschemewithsegmentationoptimization}

A typical bacterial genome contains on the order of $N \sim 10^6$
bases, within which we can justify at most $M \sim 10^3$ statistically
stationary segments.  The segmentation problem thus involves optimally
placing $M$ domain walls $\{i_1, \dots, i_m, \dots, i_M\}$.  If we
place no restrictions on where $i_m$ can be, apart from the fact that
it must be an integer between 1 and $N$, our high-dimensional
optimization problem lives in an enormous maximal search space 
\begin{equation}
\mathcal{M} = 
\{ (i_1, \dots, i_m, \dots, i_M) | 1 \leq i_1, \dots, i_m, \dots, i_M \leq N \}
\end{equation}
consisting of $N^M \sim 10^{6000}$ points.  The actual search space
is much smaller, because we must satisfy the constraint $1 < i_1 < i_2
< \cdots < i_M < N$, and smaller still if we demand that $i_{m+1} -
i_m$, for statistical reasons, must be larger than some minimum
separation.  But even this realistic search space is huge, and there
is no good global algorithm for moving the $M$ domain walls
simultaneously from an initial point in the search space, whatever
objective function (e.g. $M$-segment sequence likelihood, net 
strengths of the domain walls, net deviation of segments from
stationarity) we choose for the optimization problem.

Clearly, some complexity reduction strategy is needed to solve this
high-dimensional optimization problem.  In the recursive
Jensen-Shannon segmentation scheme of Bernaola-Galv\'an \emph{et al.}
\cite{BernaolaGalvan1996PhysicalReviewE53p5181,
RomanRoldan1998PhysicalReviewLetters80p1344}, the complexity of the
full problem is reduced by breaking the search for $M$ domain walls
into $\log M$ stages.  At each stage of the search, the computational
complexity is further reduced by the restriction that there shall be
at most one new domain wall between every existing pair of adjacent
domain walls.  We review this recursive segmentation scheme in
Sec.~\ref{subsection:needsegmentationoptimization}, and explain that
while this method is conceptually appealing, the mean-field analysis
in Sec.~\ref{subsection:meanfieldwindowlessspectrum} tells us that the
segmentation obtained at each stage of the recursion is not optimal.
We then describe in Sec.~\ref{subsection:updatescheme} two local
optimization schemes, in which a single domain wall is moved each
time, and discuss how these schemes that can be incorporated into the
recursive Jensen-Shannon segmentation framework.

\subsection{The Need for Segmentation Optimization}
\label{subsection:needsegmentationoptimization}

While not explicitly stated in their formulation, Bernaola-Galv\'an
\emph{et al.} assumed that the segment structures of genomic sequences
are organized hierarchically, i.e.~strongly distinct segments are
long, and contain less distinct segments that are shorter, which in
turn contain even less distinct segments that are shorter still.
Within this hierarchical picture of the genomic sequence, the
recursive segmentation algorithm listed below:
\begin{enumerate}

\item for a given sequence $\mathbf{x} = x_1 x_2 \cdots x_N$, compute
the Jensen-Shannon divergence spectrum $\Delta(i)$ at each cursor
position $1 < i < N$, and cut $\mathbf{x}$ at the \emph{divergence
maximum} $i^*$, where $\Delta(i^*) = \max_i \Delta(i)$, into two
segments;

\item cut each segment at its divergence maximum into two subsegments,
and continue doing so recursively;

\item whenever a new cut is made, check whether a \emph{termination
criterion} is met,

\end{enumerate}
is expected to first discover the strongest domain walls, and then
progressively weaker domain walls.  The segmentation of a given
sequence will in this way be progressively refined, by adding new
domain walls to the existing set of domain walls at every recursion,
until a \emph{terminal segmentation} corresponding to a prescribed
termination criterion is obtained.

From the simple analysis presented in
Sec.~\ref{subsection:meanfieldwindowlessspectrum}, we learned that the
situation is not quite so simple.  Indeed, within the mean-field
picture, all domain walls will be discovered if we allow the recursive
segmentation to go to completion.  However, the effective strengths of
the domain walls change with each recursion, with some strong domain
walls becoming weak, and other weak domain walls becoming strong.
Consequently, the domain walls are not discovered in order of their
true strengths, and an incomplete segmentation may pick up weak
domain walls, but miss stronger ones.  For real discrete sequences
subject to local statistical fluctuations, we can never be sure by the
hypothesis testing or model selection processes that we have exhausted
all domain walls, so getting incomplete segmentations is a very real
worry.  In their early paper
\cite{RomanRoldan1998PhysicalReviewLetters80p1344},  Rom\'an-Rold\'an
\emph{et al.} noticed the domain wall strengths changing as the
segmentation is recursively refined, and their solution was to reject
a new cut if it causes the statistical significance of existing domain
walls to fall below the confidence limit.  This \emph{ad hoc}
modification of the termination criterion was questioned by Li in Ref.
\citeonline{Li2001Gene276p57}.

Since the Jensen-Shannon divergence tells us how much better a
segmented model fits the observed sequence compared to an unsegmented
model, a strong domain wall improves the sequence likelihood more than
a weak domain wall does.  By retaining strong domain walls that have
become weak in the segmentation, and rejecting new strong cuts that
would cause existing domain walls to weaken further, we will not be
picking the best $M$-segment model at each recursion.  The final set
of recursively determined domain walls is therefore likely to miss
some strong domain walls, no matter how sophisticated the hypothesis
testing on the statistical significance of each new cut, and how much
care is taken to ensure that existing domain walls remain significant.  

Ultimately, we want our segmentation, if incomplete, to consist of the
strongest possible set of domain walls.  This can be obtained, if we
trade weak domain walls for stronger ones by moving existing domain
walls from weaker to stronger positions, or remove weak domain walls,
and let the segmentation scheme find stronger replacements in the next
recursion step.  Realizing this, Li proposed the use of branch-merging
algorithms developed in the field of recursive partitioning and
tree-based methods, revisiting all domain walls to see whether their
removal will increase the statistical significance of the
segmentation.  However, this was not done in Ref.
\citeonline{Li2001Gene276p57}, where this was proposed, nor in any
papers to date, as far we know.  Instead of removing weak domain
walls, we will present in Sec.~\ref{subsection:updatescheme} two
local optimization schemes for moving domain walls from weaker to
stronger positions, in the spirit of relaxation methods used in
optimization and numerical solution of partial differential equations.

\subsection{Segmentation Optimization Schemes}
\label{subsection:updatescheme}

\begin{figure}[hbtp]
\centering
\includegraphics[scale=0.5]{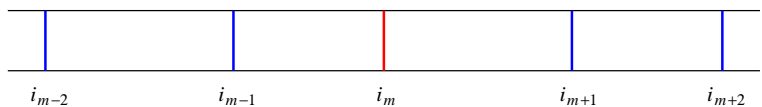}
\caption{The supersegment $(i_{m-2}, i_{m+2})$ and its four component
segments, $(i_{m-2}, i_{m-1})$, $(i_{m-1}, i_m)$, $(i_m, i_{m+1})$,
and $(i_{m+1}, i_{m+2})$, affected by the moving of the domain wall
$i_m$.}
\label{figure:moveim}
\end{figure}

From Fig. \ref{figure:moveim}, we see that moving the domain wall
$i_m$ changes the statistics of the two segments, $(i_{m-1}, i_m)$ and
$(i_m, i_{m+1})$, and as a result, the strengths of three domain
walls, $i_{m-1}$, $i_m$, and $i_{m+1}$, change.  Otherwise, the effect
of moving $i_m$ is entirely contained within the \emph{supersegment}
$(i_{m-2}, i_{m+2})$.  We can of course compute the likelihood of the
supersegment $(i_{m-2}, i_{m+2})$ directly, for each $i_m$, to
determine the position $i_m^*$ optimizing this supersegment
likelihood.  However, such a supersegment likelihood alone will not
tell us whether the domain wall $i_m^*$ is statistically significant.
This is why we fall back on the Jensen-Shannon divergence, to justify
selecting the four-segment model of $(i_{m-2}, i_{m+2})$, as opposed
to a null model containing fewer domain walls.  When we move only
$i_m$, there are two such null models, which suggest two ways to go
about optimizing $i_m$.  We call these the \emph{first-order
segmentation optimization scheme}, and the \emph{second-order
segmentation optimization scheme}.  No higher-order optimization
schemes are possible, if we allow only one domain wall to move at a
time.

In the first-order segmentation optimization scheme, we compute the
Jensen-Shannon divergence spectrum $\Delta(i)$ of the supersegment
$(i_{m-1}, i_{m+1})$, and move $i_m$ to the divergence maximum of this
supersegment.  Here, the four-segment model is compared against a
three-segment model, with segments $(i_{m-2}, i_{m-1})$, $(i_{m-1},
i_{m+1})$, and $(i_{m+1}, i_{m+2})$.  This is a natural comparison to
make, but not the only one.  In Ref.
\citeonline{Li2002CompChem26p491}, Li \emph{et al.} reported an 
experiment to determine the replication origins and replication
termini of circular bacterial genomes.  They cut a circular genome at
an arbitrary position $i_1$ to make it into a linear sequence, and
then determine the divergence maximum $i_2$ of this linear sequence.
By varying $i_1$, they found that $i_2$ would remain essentially
constant, before changing abruptly to another almost constant value.
Li \emph{et al.} identified these two stable divergence maxima with the
replication origin and replication terminus of the circular bacterial
genome.  They also noticed that $\Delta(i_2)$ reaches a maximum when
$i_1$ is either at the replication origin or the replication terminus.

This observation can be stated more generally as follows: given four
fixed domain walls $i_1$, $i_2$, $i_4$, and $i_5$, and a variable
domain wall $i_3$, such that $i_1 < i_2 < i_3 < i_4 < i_5$, the
sum $\Delta(i_2) + \Delta(i_4)$ of Jensen-Shannon divergences
comparing the four-segment model with segments $(i_1, i_2)$, $(i_2,
i_3)$, $(i_3, i_4)$, and $(i_4, i_5)$ against the two-segment model
with segments $(i_1, i_3)$ and $(i_3, i_5)$ is maximum when $i_3$ is
at a true domain wall position $i_3^*$.  Therefore, in the
second-order segmentation optimization scheme, we would compute
$\Delta(i_{m-1}) + \Delta(i_{m+1})$ over the supersegment $(i_{m-2},
i_{m+2})$ as we vary $i_m$, and move $i_m$ to the point
maximizing the sum of divergences.  As we can see from the null models
used, the two segmentation optimization schemes are not equivalent.
Nevertheless, we find in pilot numerical studies on real bacterial
genomes that the segmentations they produce are always in strong
agreement.

Because these optimization schemes are local updates, they must be
applied in turn to the domain walls $\{i_m\}_{m=1}^M$ in the
segmentation.  Clearly, if we move $i_{m+1}$ after moving $i_m$, there
will be no guarantee that $i_m$ remains optimal within either schemes.
Therefore, the optimization of the $M$ domain walls must be iterated,
until a fixed-point segmentation is obtained.  Apart from the need to
iterate the local moves, both optimization schemes can be implemented
serially, or in parallel, right after new cuts have been made in Step
1, and before further cuts are made in Step 2 of the recursive
segmentation.  Specifically, for the first-order optimization scheme,
which is simpler to implement and thus used exclusively for our
segmentation studies, we can optimize all the even domain walls
simultaneously, before optimizing all the odd domain walls
simultaneously.

\section{Segmentation Termination Condition}
\label{section:segmentationterminationcondition}

In the recursive segmentation scheme described in
Sec.~\ref{section:recursiveschemewithsegmentationoptimization}, the
number of statistically significant domain walls is known only after
segmentation is terminated everywhere in the sequence.  For an
existing segment, this involves making a decision to stop further
refinement, based on some termination criterion derived within a
hypothesis testing framework
\cite{BernaolaGalvan1996PhysicalReviewE53p5181,
RomanRoldan1998PhysicalReviewLetters80p1344}, or a model selection
framework \cite{Li2001PhysicalReviewLetters86p5815,
Li2001ProcRECOMB01p204}.  The chief shortcoming of these termination
criteria is that their common assumption that the appropriate null
model is that of a statistically stationary sequence.  As observed by
Fickett \emph{et al.}, there is generally less local homogenuity than
necessary in the sequence statistics to justify such a null model
\cite{Fickett1992Genomics13p1056}.

\begin{figure}[htbp]
\centering
\includegraphics[scale=0.45,clip=true]{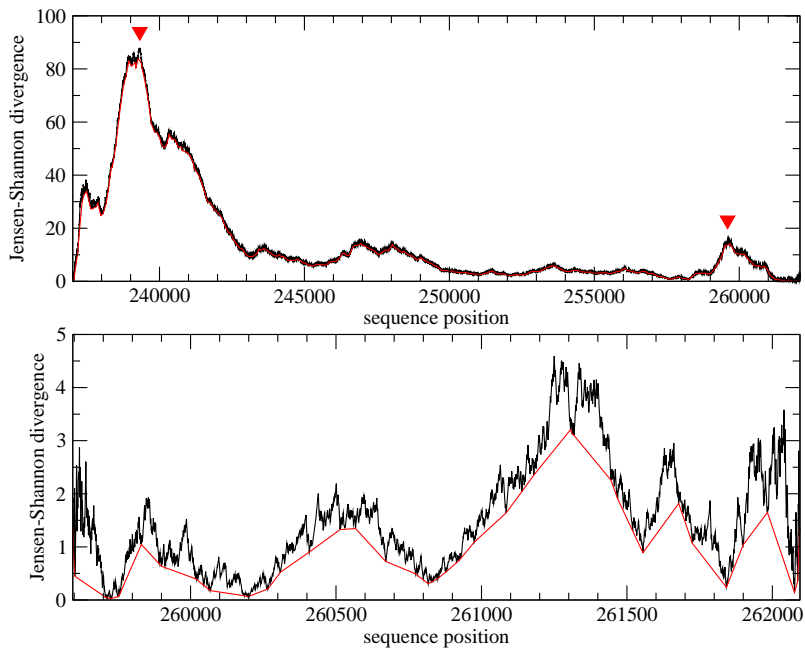}
\caption{(Top) The $K = 0$ Jensen-Shannon divergence spectrum (black)
for the interval $(237007, 262095)$, bound by the tRNAs \emph{aspV}
$(236931, 237007)$ and \emph{thrW} $(262095, 262170)$ on the \emph{E.
coli} K-12 MG1655 genome.  The two peaks highlighted are $i = 239320$
and $i = 259595$.  (Bottom) The $K = 0$ Jensen-Shannon divergence
spectrum (black) for the interval $(259595, 262095)$ of the \emph{E.
coli} K-12 MG1655 genome.  While $(259595, 262095)$ is a subinterval
of $(237007, 262095)$, the bottom divergence spectrum is not a blow up
version of the same region in the top divergence spectrum.  In both
plots, the red spectra are derived from the respective black spectra
by approximate coarse graining, assuming a minimum statistically
reliable segment length of $n = 128$.}
\label{figure:tocutornottocut}
\end{figure}

Indeed, we saw in Fig. \ref{figure:JS10} that the mean-field
divergence spectrum of the ten-segment sequence looks nothing like
that of a two-segment sequence.  Nevertheless, we should make cuts to
this ten-segment sequence, and in the mean-field limit, keep doing so
until the divergence spectra of the segments vanish identically.  For
real sequences, we find that as we segment the sequence at a finer and
finer scale, the divergence spectrum looks less and less like that
coming from a one-segment, or two-segment sequence.  This suggests
that we probably should not be doing hypothesis testing, or model
selection between one-segment and two-segment models of a given
sequence, but look at properties intrinsic to the statistical
fluctuations.  For example, in Fig.~\ref{figure:tocutornottocut}, we
show the divergence spectra of a long sequence (the interval
$(237007, 262095)$ in the \emph{E. coli} K-12 MG1655 genome, bound by
two tRNAs), which we clearly should segment, and a short sequence (the
interval $(259595, 262095)$ in the \emph{E. coli} K-12 MG1655 genome,
consisting of the \emph{proAB} operon
\cite{Salgado2006NucleicAcidsResearch34pD394}), which we are inclined
not to further segment.  The key feature that led us to our intuitive
decision is the strength of the peak relative to the typical
statistical fluctuations.

\begin{figure}[htbp]
\centering
\includegraphics[scale=0.45,clip=true]{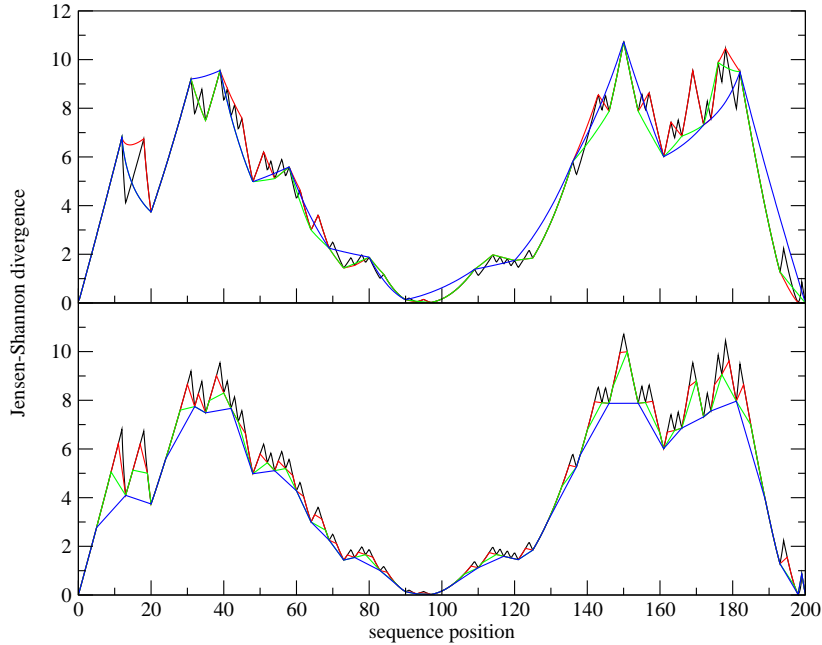}
\caption{(Top) A series of $K = 0$ mean-field Jensen-Shannon
divergence spectra for an artificial binary sequence of length $N =
200$.  At the finest scale, shown as the black divergence spectrum,
the sequence is allowed to contain unit-length segments.  When we
require the sequence to contain segments no shorter than $n = 2, 4,
8$, we obtain the red, green, and blue divergence spectra
respectively.  (Bottom) The $K = 0$ Jensen-Shannon divergence spectrum
(black) for an artificial binary sequence of length $N = 200$, and the
divergence spectra obtained from coarse graining approximately at
length scales $n = 2$ (red), $n = 4$ (green), and $n = 8$ (blue).}
\label{figure:JSseries}
\end{figure}

But what distinguishes statistical fluctuations from a genuine
statistical trend, for example, that lead to the divergence peak at $i
= 259595$ within the interval $(237007, 262095)$ (see top plot in
Fig.~\ref{figure:tocutornottocut})?  To answer this question, let us
consider a 200-segment binary sequence, where each segment is of
length one, and has unit count for either `0' or `1'.  In this binary
sequence, we find long strings of `0's and `1's, as well as regions
where the sequence alternates rapidly between `0's and `1's.  These
correspond to smooth and spiky regions in the mean-field divergence
spectrum, shown as the black curve in the top plot of
Fig.~\ref{figure:JSseries}, respectively.  Based on the mean-field
divergence spectrum, we can think of the sequence as comprising $M <
200$ long and short segments, the shortest of which are of unit
length.  Let us call the divergence spectrum of a sequence containing
unit-length segments the \emph{raw divergence spectrum} $\Delta(i)$.

Of course, it makes no statistical sense to talk about unit-length
segments, but let us pretend that it is perfectly alright to have
segments with length $n = 2$.  This being the case, the $n = 1$
segments must be absorbed into its longer flanking segments, or merged
amongst themselves, to give segments that are at least $n = 2$ long.
This is the sequence segmentation problem in a different guise, so
there are no simple solutions.  There are, however, many statistically
reasonable ways to perform this $n = 1 \to n = 2$ coarse graining, and
one of them gives rise to the red mean-field divergence spectrum shown
in Fig.~\ref{figure:JSseries}.  We can repeat this coarse-graining
procedure, coarse graining $n = 2$ segments to obtain segments at
least $n = 4$ long, giving the green mean-field divergence spectrum,
and then again, coarse graining $n = 4$ segments to to obtain segments
at least $n = 8$ long, giving the blue mean-field divergence spectrum,
both of which are shown in the top plot of Fig. \ref{figure:JSseries}.

As we can see, by making the shortest segments in the sequence longer
and longer, the mean-field divergence spectrum becomes smoother and
smoother.  Once we have achieved the desired minimum segment length
$n$, we can compare the coarse-grained divergence spectrum
$\bar{\Delta}(i)$ to the raw divergence spectrum $\Delta(i)$.  The
strength of the statistical fluctuations in the sequence can then be
quantified by the integrated absolute difference between the two
divergence spectra, 
\begin{equation}
\delta A = \int_0^N di\, \left|\bar{\Delta}(i) - \Delta(i)\right|,
\end{equation}
for this given final $n$.  To decide whether we should segment the
sequence, we compare this integrated statistical fluctuation $\delta
A$ against the total area 
\begin{equation}
A = \int_0^N di\, \Delta(i)
\end{equation}
under the raw divergence spectrum.  If the ratio $\delta A/A$ is
small, like for the top divergence spectrum in
Fig.~\ref{figure:tocutornottocut}, a cut placed at the divergence
maximum will be statistically significant.  If $\delta A/A$ is large,
like for the bottom divergence spectrum in
Fig.~\ref{figure:tocutornottocut}, a cut placed at the divergence
maximum will not be statistically significant.

In practice, coarse graining a segmentation at scale $n$ (meaning that
the shortest segments are at least $n$ long) to give a segmentation at
scale $2n$ is a time-consuming optimization problem, so we devise an
approximate form of the test statistic $\delta A/A$ to quantify the
strength of statistical fluctuations in the divergence spectrum.  This
approximate test statistic is constructed for a discrete sequence as
follows:
\begin{enumerate}

\item for a given set of sequence positions $\{i_r\}_{r=1}^M$ and
divergences $\{\Delta_r = \Delta(i_r)\}_{r=1}^M$, partition the
sequence positions into a set of \emph{convex} points, satisfying
\begin{equation}
\Delta_{r} \leq 
\frac{i_{r} - i_{r-1}}{i_{r+1} - i_{r-1}} \Delta_{r-1} + 
\frac{i_{r+1} - i_{r}}{i_{r+1} - i_{r-1}}\Delta_{r+1},
\end{equation}
and a set of \emph{concave} points, which fail the above convexity
condition;

\item iterate through the set of concave points, and retain the
concave point $i_r$ if $i_r - i_{r-1} > n$ and $i_{r+1} - i_r > n$.
Otherwise, we reject $i_r$, and all convex points within $n$ of it.
The set of retained concave points, together with convex points that
are not rejected, form a coarse-grained set of domain walls that are
at least $n$ apart from each other, and the divergences at these
points give a coarse-grained divergence spectrum; 

\item multiply $n$ by two, and if the new length scale is less than
the desired final length scale $n_c$, repeat steps 1 and 2.
Otherwise, stop the coarse graining process.

\end{enumerate}
In the bottom plot of Fig.~\ref{figure:JSseries}, we show the
divergence spectra obtained from coarse graining approximately at
length scales $n = 2, 4, 8$.

For real genomes, we require this approximate coarse graining to
proceed until $n_c = 128 \cdot 4^{K^*}$ for a quaternary sequence
whose optimal Markov-chain order is $K^*$.  This is to ensure that
each transition count would be at least 20--30, so that the
maximum-likelihood transition probabilities can be reliably estimated.
Going back to Fig.~\ref{figure:tocutornottocut}, we see that the
$n_c = 128$ approximate coarse-grained divergence spectrum (red curve)
for the long interval $(237007, 262095)$ is very close to its raw
divergence spectrum (black curve), whereas for the short interval
$(259595, 262095)$, the $n_c = 128$ approximate coarse-grained
divergence spectrum (red curve) is significantly different from its
raw divergence spectrum (black curve).  If we compute $\delta A$ using
the approximate coarse-grained divergence spectra $\bar{\Delta}(i)$
for the two intervals, we will find a small $\delta A/A$ for the long
interval, and a large $\delta A/A$ for the short interval.

Based on the above discussions, we propose to use
\begin{equation}
\frac{\delta A}{A} > \left(\frac{\delta A}{A}\right)^*
\end{equation}
as an alternative termination criterion, where $\left(\delta
A/A\right)^*$ is a user prescribed tolerance.  The test statistic
$\delta A/A$ requires no prior assumption on how many segments to
partition the sequence into, but instead requires that the shortest
segments used to model the sequence must yield adequate statistics to
reliably estimate the model parameters.  We believe that $\delta A/A$,
which measures quantitatively the relative strength of statistical
fluctuations up to some cutoff length scale $n_c$, is a better test
statistic to use as the termination criterion, as there is no bias
towards any particular segment model of the given sequence.  Based on
extensive experimenting, we find that $\left(\delta A/A\right)^* =
0.30$ will stop the recursive segmentation just before individual
genes are segmented.

\section{Application to Real Genomes}
\label{subsection:recursivegenome}

\begin{figure}[htbp]
\centering
\includegraphics[scale=0.75]{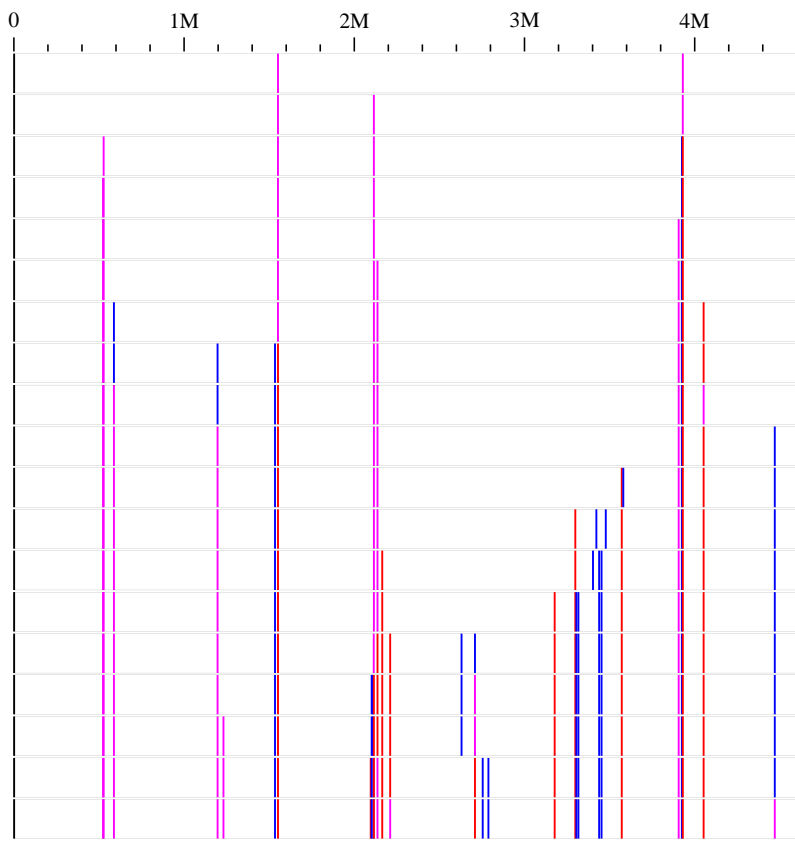}
\caption{Series of recursive Jensen-Shannon segmentations of the
\emph{E. coli} K-12 MG1655 genome ($N = 4639675$ bp), with and without
segmentation optimization, for (top to bottom) $2 \leq M \leq 20$
domain walls.  In this figure, unoptimized domain walls are shown in
red, optimized domain walls are shown in blue, and unoptimized and
optimized domain walls agreeing to within 2000 bp of each other are
shown in magenta.  The two domain walls that appear in the $M = 2$
optimized segmentation are close to the replication origin and
replication terminus, and remain close to where they were first
discovered as the recursive segmentation progresses.}
\label{figure:hierarchyofrecursivesegmentations}
\end{figure}

In the original recursive segmentation algorithm proposed by
Bernaola-Galvan \emph{et al.}
\cite{BernaolaGalvan1996PhysicalReviewE53p5181,
RomanRoldan1998PhysicalReviewLetters80p1344}, it does not matter how
many new cuts are added to the sequence at each recursion step --- we
get the same terminal segmentation for the same termination criterion.
However, if we optimize the segmentation before new cuts are added,
then the terminal segmentation depends very sensitively on the
termination criterion.  To benchmark the performance of the recursive
Jensen-Shannon segmentation scheme, with and without segmentation
optimization, we add only one new cut --- the strongest of all new
cuts possible --- every recursion step.  A series of recursive
segmentations (with and without segmentation optimization) obtained
this way is shown in Fig.
\ref{figure:hierarchyofrecursivesegmentations} for the \emph{E. coli}
K-12 MG1655 genome, for $2 \leq M \leq 20$ domain walls in the
sequence.

From Fig. \ref{figure:hierarchyofrecursivesegmentations}, we find that
in contrast to the truly hierarchical unoptimized recursive
segmentations, the optimized recursive segmentations of \emph{E. coli}
K-12 MG1655 are almost hierarchical, i.e. few domain walls in 
segmentations containing more domain walls are shifted relative to 
segmentations containing fewer domain walls.  Those few domain walls
which do get shifted, however, can be shifted a lot.  For example, we
see the domain wall $i_{10} = 4051637$ in the optimized segmentation
with $M = 10$ domain walls get shifted to $i_{10} = 4469701$ in the
optimized segmentation with $M = 11$ domain walls ($\delta i_{10} =
+418064$).  Another example is $i_7 = 2135183$ in the optimized
segmentation with $M = 15$ domain walls shifted to $i_7 = 2629043$ in
the optimized segmentation with $M = 16$ domain walls ($\delta i_7 =
+493860$).  We also observed that domain walls `lost' as a result of
optimization frequently reappear later in the optimized recursive segmentation.
For example, the domain wall $i_{10} = 4051637$, which
was shifted to $i_{10} = 4469701$ when a new cut was made to the
optimized segmentation with $M = 10$ domain walls, reappears in the
optimized segmentation with $M = 29$ domain walls (not shown in Fig.
\ref{figure:hierarchyofrecursivesegmentations}).  These are
manifestations of what we call the \emph{context sensitivity problem},
which we will carefully study in another paper \cite{Cheong2007CSP}.

In the unoptimized recursive segmentation scheme, a domain wall
remains at where it was first discovered, and its statistical
significance is measured solely by its strength.  However, the
strength of existing domain walls frequently change as more and more
domain walls are added to the segmentation, so it is not clear what
kind of statistical significance to attach to a domain wall that
becomes progressively weaker (or one that becomes progressively
stronger, for that matter).  In the optimized recursive segmentation
scheme, we find that there are some domain walls that remain close to
where they are first discovered, which we call \emph{stable} domain
walls, and those that do not as recursion proceeds.  Stability with
respect to segmentation refinement-optimization provides an
alternative measure of statistical significance, in the sense that at
any level of segmentation, a stable domain wall is more significant
than a domain wall that has been shifted around for the past few
recursions.  A domain wall that has been stable over a larger number
of recursions is also more significant than a domain wall that has
been stable over a smaller number of recursions.  From Fig.
\ref{figure:hierarchyofrecursivesegmentations}, we see that stable
domain walls are always discovered first by the optimized segmentation
scheme, if not simultaneously by both schemes.

\section{Conclusions}
\label{section:conclusions}

To summarize, we have in this paper presented improvements to three
different aspects of the recursive Jensen-Shannon segmentation scheme
proposed by Bernaola-Galv\'an \emph{et al}.  In
Sec.~\ref{section:generalizedJensenShannondivergences}, which touch on
the modeling aspects of recursive segmentation, we explained how
Markov chains of order $K > 0$, producing better fits of the higher
order statistics with fewer independent parameters, are better models
of the segments within a heterogeneous genomic sequence, compared to
Bernoulli chains over the quaternary or extended alphabets.  We then
wrote down a generalized Jensen-Shannon divergence, given in
Eq.~\eqref{equation:JensenShannondivergence}, which can be used as an
entropic measure to distinguish between two different Markov chains of
the same order.  Moving on, we described how to select a maximum
Markov-chain order for the segmentation of a given sequence, by
minimizing its Bayes information criterion (BIC), and how to optimize
the position of the domain wall between two segments with different
maximum Markov-chain orders.

Next, in Sec.~\ref{section:meanfieldanalysis}, we described how the
spatial distribution of rare transitions along the sequence give rise
to local fluctuations in the Jensen-Shannon divergence spectrum,
confusing our intuitive picture of how recursive segmentation proceeds
in a heterogeneous sequence.  We argued that a mean-field analysis,
which we then undertook, would be useful in better understanding
recursive segmentation.  We found that in the mean-field
Jensen-Shannon divergence spectrum, true domain walls can appear
either as peaks or kinks, or even have vanishing divergences.  We
showed that all true domain walls will eventually be discovered, if we
allow the recursive segmentation to go to completion, but these will
in general not be discovered in the order of their true strengths.
Consequently, an incomplete segmentation will generically be
suboptimal, because it contains weak domain walls that are discovered
ahead of stronger ones, which are thus left out of the segmentation,
and we argued in
Sec.~\ref{section:recursiveschemewithsegmentationoptimization} that
there is a need to optimize the segmentation obtained at every stage
of the recursive entropic segmentation.  We then proposed two local
optimization schemes to move domain walls from weaker to stronger
positions, which improves the statistical significance of an
incomplete segmentation.  

In Sec.~\ref{section:segmentationterminationcondition}, we improved
on the statistical testing aspect of the recursive Jensen-Shannon
segmentation scheme by proposing a new termination criterion.  We
motivated the test statistic associated with this new termination
criterion by considering the problem of recursively coarse graining an
initial segmentation containing very short segments to produce a final
segmentation containing no segments shorter than a cutoff length scale
$n_c$.  We devised an algorithm for performing this recursive coarse
graining approximately, and argued that the fraction difference
$\delta A/A$ in areas under the raw divergence spectrum of the initial
segmentation and the smoothed divergence spectrum of the final
segmentation, can be used as the test statistic for terminating
recursive segmentation.  The principal advantage of using this new
termination criterion, which is a quantitative measure of the strength
of sub-$n_c$ statistical fluctuations relative to the maximum
divergence in the raw divergence spectrum, is that it requires no
prior knowledge of how many segments to partition the sequence into,
and thus has no need to assume a single-segment null model for the
sequence.

Finally, in Sec.~\ref{subsection:recursivegenome}, we performed a
benchmark study on the genome of the model bacterium \emph{E. coli}
K-12 MG1655, comparing the recursive Jensen-Shannon segmentation
schemes with and without optimization.  The optimized segmentations
obtained at different stages of the recursion were found to be not
perfectly hierarchical, because of the context sensitivity problem
which we will investigate in another paper \cite{Cheong2007CSP}.  At
the coarse levels of segmentations examined, we found the recursive
segmentation schemes with and without segmentation agreeing on a set
of stable domain walls, but not on the order these are discovered.  We
argued that the stability of a domain wall with respect to optimized
recursive segmentation is an alternative measure of its statistical
significance, and found that these stable domain walls are always
discovered first by the optimized segmentation scheme, if not
simultaneously by both schemes.

Besides the context sensitivity problem which we realize plagues all
segmentation schemes \cite{Cheong2007CSP}, the innovations and
discoveries in this paper also inspired other studies.  In a future
paper \cite{Cheong2007CTS}, we will report a clustering analysis on
the terminal genomic segments, obtained using the optimized recursive
Jensen-Shannon segmentation scheme with the new termination criterion,
of a bacterial plant pathogen.  Based on this clustering analysis on a
single bacterial genome, we will construct its statistical genetic
backbone, and identify possible horizontally transferred genetic
islands.  We will then extend the clustering analysis for comparative
studies of several closely related bacteria, where we will detect
syntenic regions, and identify a phylogenetic backbone.  We would like
to suggest that, even though the optimized recursive segmentation
scheme presented in this paper is formulated as a method for
biological sequence segmentation, it can also be applied in other
engineering segmentation problems (for example, in image segmentation,
and change point detection in noisy time series).

\section*{Acknowledgment}

SAC and CRM acknowledge support from USDA Agricultural Research
Service project 1907-21000-017-05.  Computational resources of the
USDA-ARS plant pathogen systems biology group and the Cornell Theory
Center assisted this research.

\begin{biography}[%
	{\includegraphics[%
		width=1in,height=1.25in,clip,keepaspectratio]{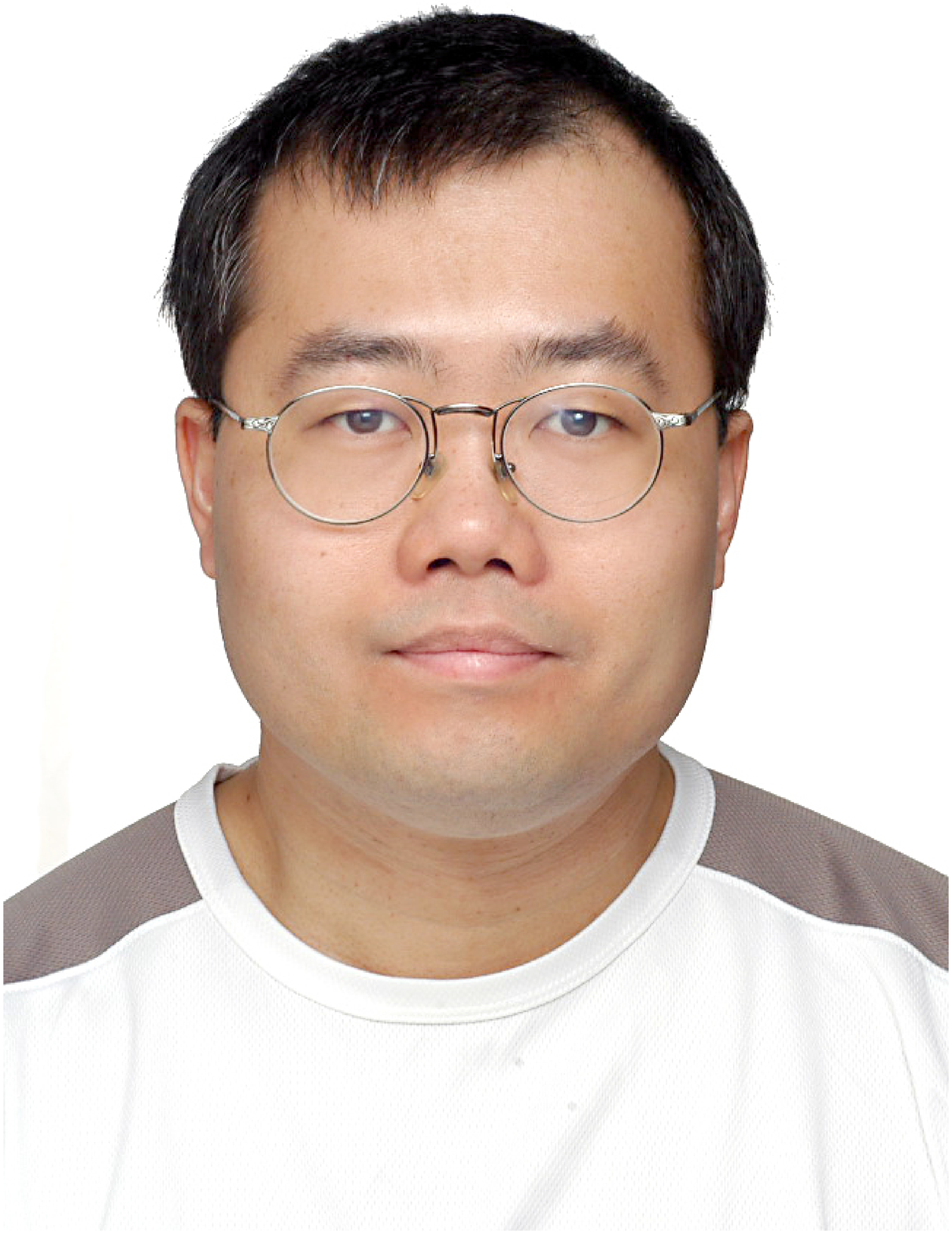}}]%
		{Siew-Ann Cheong}
received his BSc with first-class honors in physics in 1997 from the
National University of Singapore, and his PhD in physics in 2006 from
Cornell University.  He was a postdoctoral associate with
the Cornell Theory Center between 2006 to 2007, and is now an
assistant professor in the Division of Physics and Applied Physics,
School of Physical and Mathematical Sciences, Nanyang Technological
University, Singapore.  His research interest is on the use of
statistical methods to discover meso- and macro-scale structures in
bacterial genomes.
\end{biography}

\begin{biography}[%
	{\includegraphics[%
		width=1in,height=1.25in,clip,keepaspectratio]{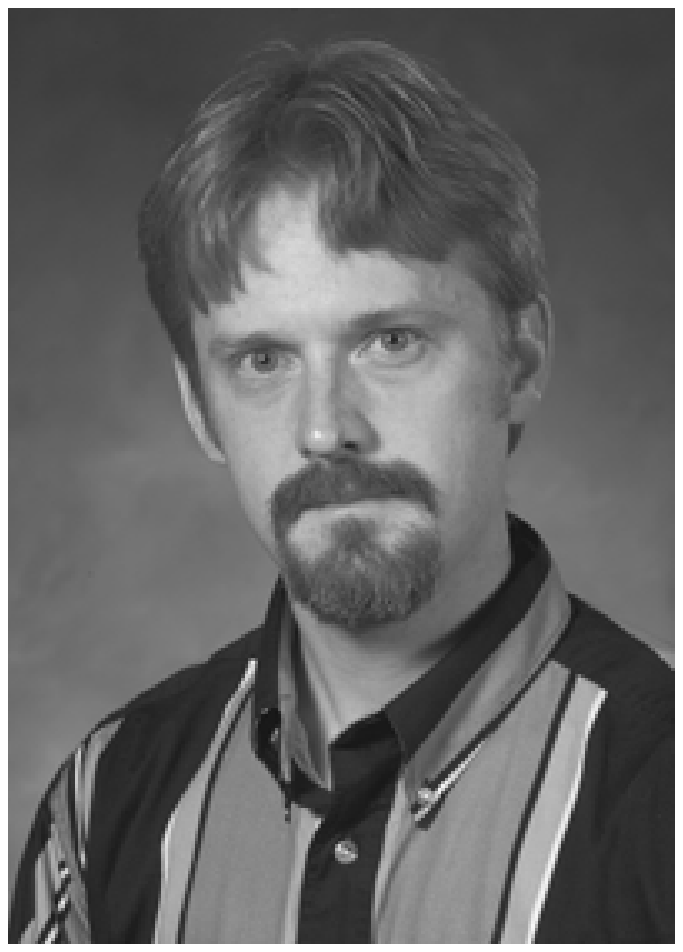}}]%
		{Paul Stodghill}
received the B.A. degree in mathematics and computer
science from Dickinson College, Carlisle, PA, in 1988 and the
Ph.D. degree from the Department of Computer Science at Cornell
University, Ithaca, NY, in 1997.  He is currently a computational
biologist for the Agricultural Research Service of the US
Department of Agriculture.
\end{biography}

\begin{biography}[%
	{}]%
		{David J. Schneider}
is a computational biologist for the Agricultural Research Service of
the US Department of Agriculture.
\end{biography}

\begin{biography}[%
	{\includegraphics[%
		width=1in,height=1.25in,clip,keepaspectratio]{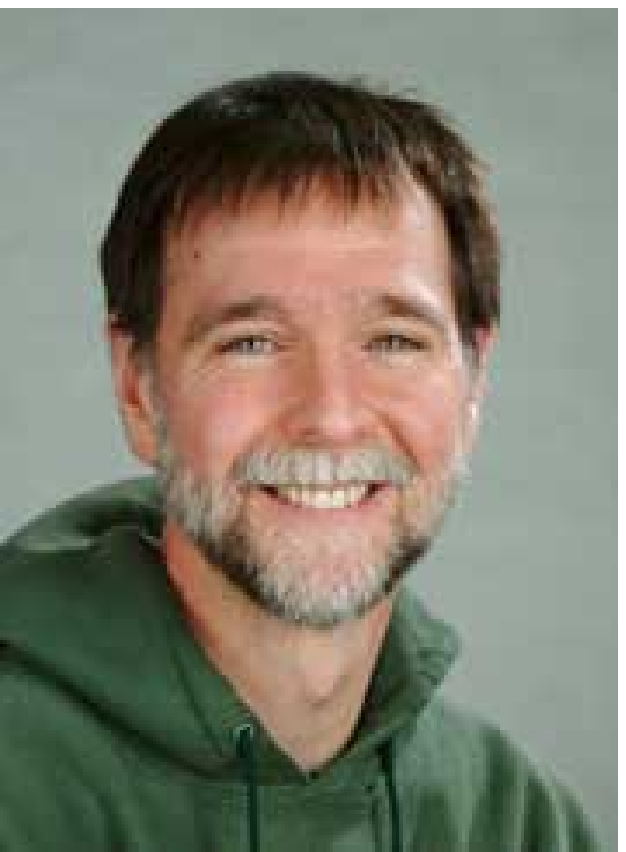}}]%
		{Samuel W. Cartinhour}
is the USDA-ARS Courtesy Professor in the Department of Plant
Pathology, Cornell University.  His research interest is in employing
computational and laboratory experiments to understand gene
regulation, and in particular, pathogenesis,
in bacterial plant pathogen \emph{Pseudomonas syringae}.
\end{biography}

\begin{biography}[%
	{\includegraphics[%
		width=1in,height=1.25in,clip,keepaspectratio]{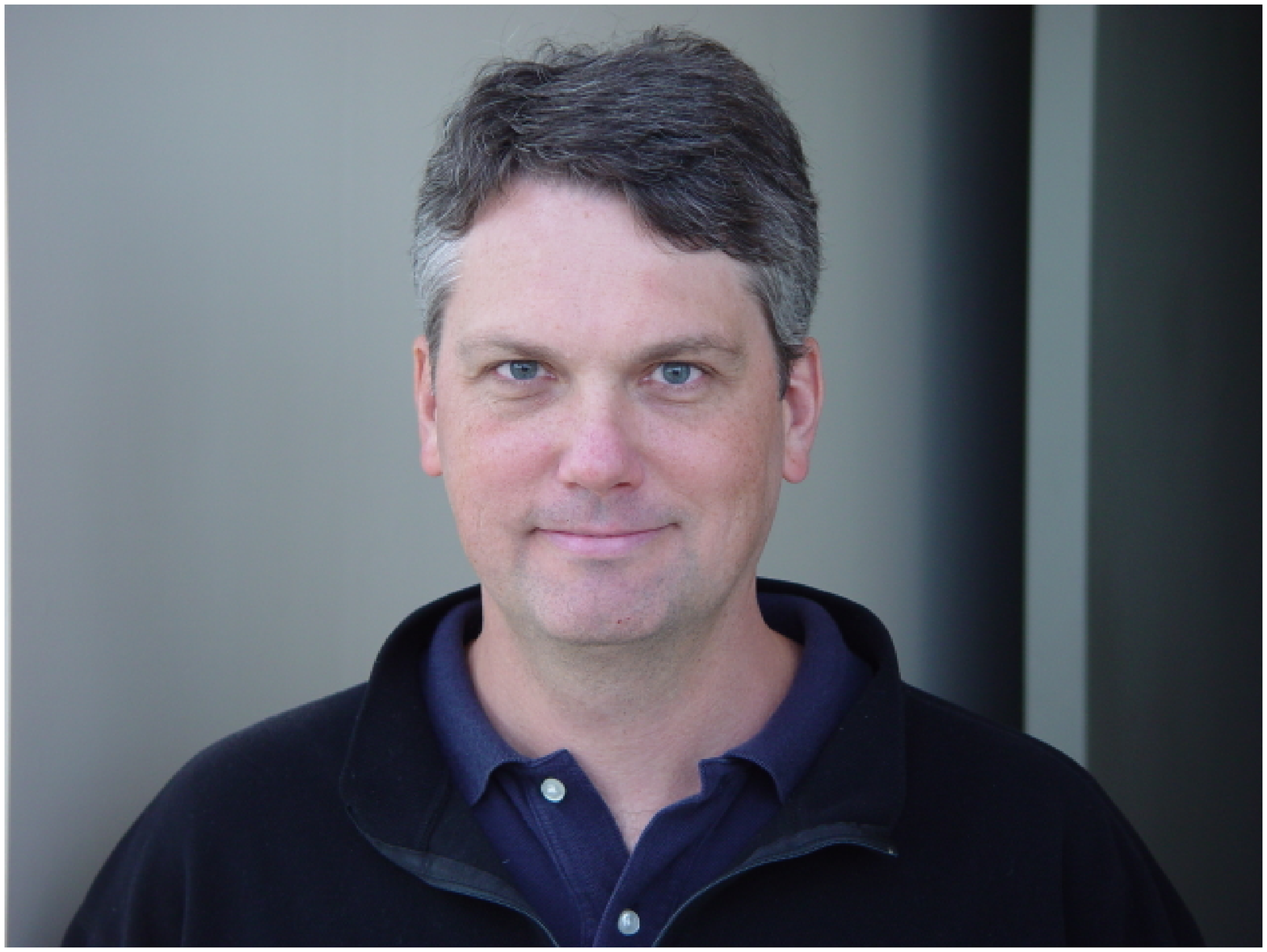}}]%
		{Christopher R. Myers}
received a B.A. in History from Yale University, and a Ph.D. in
physics from Cornell University.  His research interests snake along
interdisciplinary boundaries connecting physics, biology, and computer
science, with particular emphasis on the emergence of information
processing in cellular regulation and signaling, and on the functional
organization of complex biological networks.
\end{biography}

\end{document}